\documentstyle[12pt]{article}

\begin{document}
\title{ Casimir stress for concentric spheres in de Sitter space}
\author{
M.R.Setare\footnote{E-mail:Mreza@physics.sharif.ac.ir}
\\ Department of Physics,Sharif University of Technology,
Tehran-Iran} \maketitle

\begin{abstract}
The Casimir stress on two concentric spherical shell in de Sitter
background for massless scalar field is calculated. The scalar
field satisfies Dirichlet boundary conditions on the spheres. The
metric is written in conformally flat form to make maximum use of
Minkowski space calculations. Then the Casimir stress is
calculated for inside and outside of the shell with different
backgrounds. This model may be used to study the effect of the
Casimir stress on the dynamics of the domain wall formation in
inflationary models of early universe.
 \end{abstract}
\newpage
 \section{Introduction}
   The Casimir effect is one of the most interesting manifestations
  of nontrivial properties of the vacuum state in quantum field
  theory[1,2]. Since its first prediction by
  Casimir in 1948\cite{Casimir} this effect has been investigated for
  different fields having different boundary geometries[4-7]. The
  Casimir effect can be viewed as the polarization of
  vacuum by boundary conditions or geometry. Therefore, vacuum
  polarization induced by a gravitational field is also considered as
  Casimir effect.\\
   Casimir effect for spherical shells in the presence of the
   electromagnetic fields has been calculated several years ago\cite{{Boyer},{Schw}}. A
   recent simplifying account of it for the cases of electromagnetic and
   scalar fields with both Dirichelt and  Neumann boundary conditions on sphere
   is given in\cite{Nester}. The dependence of Casimir energy on the
   dimension of space for electromagnetic and scalar fields with Dirichlet boundary
   conditions in the presence of a spherical shell is discussed in\cite{{Milton},{Mil}}.
The Casimir energy for odd and even space dimensions and different
fields, including the spinor field, and all the possible boundary
conditions have been considered in\cite{Cog}. There it is
explicitly shown that although the Casimir energy for interior
and exterior of a spherical shell are both divergent, irrespective
of the number of space dimensions, the total Casimir energy of
the shell remains finite for the case of odd space dimensions.
Heat kernel coefficients and zeta function of the Laplace
operator on a D-dimensional ball with different boundary
conditions, both of them useful tools to calculate Casimir
energies, have been calculated in {\cite {bek1}, \cite{bek2}}.
More recently a new method have bas developed for the scalar
Casimir stress on the D-dimensional sphere, in \cite{Sahar1}. In
this reference the regularized vacuum expectation values for the
scalar field energy-momentum tensor inside and outside a
spherical shell and in the region between two concentric spheres
have been calculated. Of some interest are cases where the field
is confined to the inside of a spherical shell. This is sometimes
called the bag boundary condition. The application of Casimir
effect to the
    bag model is considered for the case of massive scalar field
    \cite{Bord} and the Dirac field \cite{Eli2}. We use the renormalization
    procedure in the above cases for our problem.\\
    Casimir effect in curved space-time has not been studied extensively.
    Casimir effect for spherical boundary in curved space-time is considered
    in\cite{{Bay},{Bay2}} where the Casimir energy for half of $S^{3}$ and $S^{2}$ with
    Dirichlet and Neumann boundary conditions for massless conformal scalar field is
  calculated analytically using all the existing methods.
  Casimir effect in the presence of a general relativistic domain
  wall is considered in \cite{Set} and a study of the relation
  between trace anomaly and the Casimir effect can be found in
  \cite{set1}. Casimir effect may  have interesting implications for the early
  universe. It has been shown, e.g., in\cite{Ant} that a closed Robertson-Walker
  space-time in which the only contribution to the  stress tensor comes from Casimir energy
   of a scalar field is excluded. In inflationary models, where the
   dynamics of bubbles may play a major role, this dynamical
   Casimir effect has not yet been taken into account. Let us
   mention that in \cite{set2} we have investigated the
   Casimir effect of a massless scalar field with Dirichlet
   boundary condition in spherical shell having different vacuua
   inside and outside which represents a bubble in early universe with
   false/true vacuum inside/outside. In this reference the sphere
   have zero thickness. In the present paper we shall extend our
   analysis to the spherical shell with nonvanishing thickness. \\
  Our aim is to calculate the Casimir stress on two concentric spherical
   shell with constant comoving radius having different vacuua inside and
    outside in de Sitter space. This configuration is corresponding to a
    spherical symmetry domain wall with thickness. We assume the inner and
    outer regions of thick shell are in $\Lambda$ vacuum corresponding to
    degenerate vacuum in domain wall configuration. This example is similar
    to the our recent study of planer cosmic domain wall
    \cite{set3}.\\
    The organization of the paper is as follows. In section two we
    consider two concentric sphere in flat space-time, we shall
    rely on the result of \cite{Sahar1} and consider vacuum
    stress on each single spheres. In section three we obtain
    renormalized Casimir energy for each single sphere in a de
    Sitter space. We use a procedure similar to that of our
    previous work \cite{set2}, then we calculate total stress on
    the sphere due to the boundary conditions and gravitational
    vacuum polarization. In last section we conclude and summarize
     the results.

  \section{Scalar Casimir stress for concentric spheres in flat space-time}
  We consider two concentric spherical shells with zero thickness and
   with radii $a$ and $b$, $a<b$. Consider now the Casimir force due to
  fluctuation of a free massless scalar field satisfying Dirichlet
  boundary conditions on the spherical shells in Minkowski
  space-time.
  The vacuum force per unit area of the inner sphere is given by \cite{Sahar1}
  \begin{equation}\label{Feq1}
  F_{a}(a,b)=F(a)-P_{a}(a,b,a),
  \end{equation}
  where $F(a)$ is the force per unit area of a single sphere with
  radius $a$, and $P_{a}(a,b,a)$ is due to the existence of the
  second sphere (interaction force). In a similar manner vacuum force
  acting on per unit area of outer sphere is
  \begin{equation}\label{Feq2}
  F_{b}(a,b)=F(b)+P_{b}(a,b,b).
  \end{equation}
  The vacuum force per unit area of a single sphere is the sum of Casimir
  forces $F_{in}$ and $F_{out}$ for inside and outside of the
  shell.
  \begin{equation}
  F(a)=F_{in}(a)+F_{out}(a),\hspace{1cm}
  F(b)=F_{in}(b)+F_{out}(b).
  \end{equation}
  Just as described in \cite{Mil,set2}, Casimir force inside and
  outside of the shell are divergent individually. In flat space
  when we add interior and exterior forces to each other,
  divergent parts will cancel each other out. Interaction forces $P_{a}(a,b,a)$ and
  $P_{b}(a,b,b)$ are finite and for Dirichlet boundary condition, are given by
  \begin{equation}\label{peq1}
  P_{a}(a,b,a)=\frac{-1}{8\pi^{2}a^{3}}\sum_{l=0}^{\infty}(2l+1)\int_{0}^{\infty}
  dz\frac{K_{\nu}^{(b)}(bz)/K_{\nu}^{(a)}(az)}
  {K_{\nu}^{(a)}(az)I_{\nu}^{(b)}(bz)-K_{\nu}^{(b)}(bz)I_{\nu}^{(a)}(az)},
  \end{equation}
 \begin{equation}\label{peq2}
  P_{b}(a,b,b)=\frac{-1}{8\pi^{2}b^{3}}\sum_{l=0}^{\infty}(2l+1)\int_{0}^{\infty}
  dz\frac{I_{\nu}^{(a)}(az)/I_{\nu}^{(b)}(bz)}
  {K_{\nu}^{(a)}(az)I_{\nu}^{(b)}(bz)-K_{\nu}^{(b)}(bz)I_{\nu}^{(a)}(az)},
  \end{equation}
  where $I_{\nu}$ and $K_{\nu}$ are modified Bessel function, and
  can be deduced from Eq.(5.15) in Ref. \cite{sahar1}
  These quantities are always negative.
  \section{Scalar Casimir stress for concentric spheres in de sitter space}
  Consider now the system of two concentric spheres in de Sitter
  space. To make the maximum use of the flat space calculation we use the
 de sitter metric in conformally flat form
  \begin{equation}\label{met}
  ds^{2}=\frac{\alpha^{2}}{\eta^{2}}[d\eta^{2}-\sum_{\imath=1}^{3}
  (dx^{\imath})^{2}],
  \end{equation}
  where $\eta$ is the conformal time:
  \begin{equation}\label{contime}
  -\infty \langle \eta \langle 0.
  \end{equation}
  The relation between parameter $\alpha$ and cosmological constant
  $\Lambda$ is given by
  \begin{equation}\label{const}
  \alpha^{2}=\frac{3}{\Lambda}.
  \end{equation}
  Under conformal transformation in four dimensions, the vacuum forces inside
  and outside for a single sphere with zero thickness are given by (see \cite{set2})
\begin{equation}\label{Fin}
 \bar{F}_{in}=\frac{\eta^{2}}{\alpha_{in}^{2}}\bar{F}_{in}=
 \frac{\Lambda_{in}\eta^{2}}{3}F_{in},
\end{equation}
\begin{equation}\label{Fout}
 \bar{F}_{out}=\frac{\eta^{2}}{\alpha_{out}^{2}}\bar{F}_{out}=
 \frac{\Lambda_{out}\eta^{2}}{3}F_{out}.
\end{equation}
In this case when we add interior and exterior forces to each
other, the Casimir force becomes divergent. To obtain $\bar{F}$,
we use a procedure similar to that of \cite{set2}. First of all
we consider that the inside region of the sphere with radius $a$
and outside region of the sphere with radius $b$ have cosmological
constant $\Lambda$ and the region between two spheres have
cosmological constant $\Lambda'$. Therefore we have \cite{set2}
\begin{equation}\label{Eeq1}
\bar E_{in}(a)=\frac{\eta^{2}\Lambda}{6a}(c_{1}+\frac{c'_{1}}
{\varepsilon}),\hspace{1cm}\bar
E_{out}(a)=\frac{\eta^{2}\Lambda'}{6a} (c_{2}-\frac{c'_{1}}
{\varepsilon})
\end{equation}
\begin{equation}\label{Eeq2}
\bar E_{in}(b)=\frac{\eta^{2}\Lambda'}{6b}(c_{1}+\frac{c'_{1}}
{\varepsilon}),\hspace{1cm}\bar
E_{out}(b)=\frac{\eta^{2}\Lambda}{6b} (c_{2}-\frac{c'_{1}}
{\varepsilon})
\end{equation}
where $c_{1}=0.008873,c_{2}=-0.003234,c'_{1}=0.001010$ and
$\varepsilon$ is cutoff parameter. As we see, each of the
energies for in-and-
outside of the shells are divergent.  \\
The energy for each single sphere is as follows:
\begin{equation}\label{rEeq01}
\bar E(a)=\frac{\eta^{2}}{6a}[(c_{1}\Lambda+c_{2}\Lambda')+
\frac{c'_{1}}{\varepsilon}(\Lambda-\Lambda')],
\end{equation}
\begin{equation}\label{rEeq02}
\bar E(b)=\frac{\eta^{2}}{6b}[(c_{1}\Lambda'+ c_{2}\Lambda)+
\frac{c'_{1}}{\varepsilon}(\Lambda'-\Lambda)].
\end{equation}
To renormalize the above Casimir energy we use a procedure
similar to that of our previous paper \cite{set2}. We consider
the classical energy for each sphere. The classical energy of a
sphere immersed in a cosmological background, as we are
considering, may be written as
\begin{equation}
E_{class}=PR^{3}+\sigma R^{2}+FR+K+\frac{h}{R}.
\end{equation}
In this way classical energy of spherical shell is determined by
above parameter, where $R$ is radius of spheres. Therefore total
energy of each sphere is given by
\begin{equation}
\tilde E(a)=\bar E(a)+E_{class}(a), \hspace{2cm}  \tilde E(b)
=\bar E(b)+E_{class}(b).
\end{equation}
The renormalization can be achieved now by shifting the parameter
$h$ in $E_{class}$ by an amount which cancels the divergent
contribution. For inner and outer spheres we have
\begin{equation}
h\rightarrow
h+\frac{\eta^{2}}{6}\frac{c'_{1}}{\varepsilon}(\Lambda'-\Lambda),
\end{equation}
\begin{equation}
h\rightarrow
h+\frac{\eta^{2}}{6}\frac{c'_{1}}{\varepsilon}(\Lambda-\Lambda').
\end{equation}
 After the
renormalization we obtain for the Casimir energies
\begin{equation}\label{rEeq1}
\bar E_{ren}(a)=\frac{\eta^{2}}{6a}(c_{1}\Lambda+c_{2}\Lambda'),
\end{equation}
\begin{equation}\label{rEeq2}
\bar E_{ren}(b)=\frac{\eta^{2}}{6b}(c_{1}\Lambda'+c_{2}\Lambda).
\end{equation}
Now we use the following relation for the stress on the shell
\begin{equation}\label{eqEF}
\frac{F}{A}=\frac{-1}{4\pi a^{2}}\frac{\partial E}{\partial a}.
\end{equation}
Then the stresses on the each single shell due to the boundary
condition are given by
\begin{equation}\label{rFeq1}
\frac {\bar F(a)}{A}=\frac{-1}{4\pi a^{2}}\frac{\partial\bar
E(a)}{\partial a}=\frac{\eta^{2}}{24\pi
a^{4}}(c_{1}\Lambda+c_{2}\Lambda'),
\end{equation}
\begin{equation}\label{rFeq2}
\frac{\bar F(b)}{B}=\frac{-1}{4\pi b^{2}}\frac{\partial\bar
E(b)}{\partial b}=\frac{\eta^{2}}{24\pi
b^{4}}(c_{1}\Lambda'+c_{2}\Lambda).
\end{equation}
Under conformal transformation, interaction forces are given by
\begin{equation}\label{inteq}
\bar
P_{a}(a,b,a)=\frac{\eta^{2}\Lambda'}{3}P_{a}(a,b,a),\hspace{1cm}
\bar P_{b}(a,b,b)=\frac{\eta^{2}\Lambda'}{3}P_{b}(a,b,b).
\end{equation}
Therefore the total stress on the spheres due to boundary
conditions are obtained
\begin{equation}\label{Ftot1}
\frac{\bar F_{a}(a,b)}{A}=\frac{\bar F(a)}{A}-\bar P_{a}(a,b,a)=
\frac{\eta^{2}}{24\pi a^{4}}(c_{1}\Lambda+c_{2}\Lambda')-
\frac{\eta^{2}\Lambda'}{3}P_{a}(a,b,a),
\end{equation}
\begin{equation}\label{Ftot2}
\frac{\bar F_{b}(a,b)}{B}=\frac{\bar F(b)}{B}+\bar P_{b}(a,b,b)=
\frac{\eta^{2}}{24\pi b^{4}}(c_{1}\Lambda'+c_{2}\Lambda)+
\frac{\eta^{2}\Lambda'}{3}P_{b}(a,b,b).
\end{equation}
 Now we consider the pure effect of vacuum polarization due to
 gravitational field without any boundary conditions. The
 renormalized stress tensor for massless scalar field in de sitter
 space is given by \cite{Birrell,Dowk}
 \begin{equation}\label{Teq}
\langle T^{\nu}_{\mu}\rangle=\frac{1}{960 \pi ^{2}\alpha
^{4}}\delta^{\nu}_{\mu}.
\end{equation}
Now, the effective pressure created by above gravitational part
is different for different parts of space-time:
\begin{equation}\label{pgeq1}
P_{in}^{g}(a)=\frac{-\Lambda^{2}}{8640\pi^{2}} \hspace{2cm}
P_{out}^{g}(a)=\frac{-\Lambda'^{2}}{8640\pi^{2}},
\end{equation}
\begin{equation}\label{pgeq2}
P_{in}^{g}(b)=\frac{-\Lambda'^{2}}{8640\pi^{2}} \hspace{2cm}
P_{out}^{g}(b)=\frac{-\Lambda^{2}}{8640\pi^{2}}.
\end{equation}
Therefore,gravitational pressures over spheres, are given by
\begin{equation}\label{pgeq}
P^{g}(a)=\frac{-1}{8640\pi^{2}}(\Lambda^{2}-\Lambda'^{2}),\hspace{1cm}
P^{g}(b)=\frac{-1}{8640\pi^{2}}(\Lambda'^{2}-\Lambda^{2}).
\end{equation}
The total stress on the spherical shells, is then given by
\begin{equation}\label{ptot1}
P^{tot}(a)=\frac{\eta^{2}}{24\pi
a^{4}}(c_{1}\Lambda+c_{2}\Lambda')-\frac{\eta^{2}\Lambda'}{3}P_{a}(a,b,a)-
\frac{1}{8640\pi^{2}}(\Lambda^{2}-\Lambda'^{2}),
\end{equation}
\begin{equation}\label{ptot2}
P^{tot}(b)=\frac{\eta^{2}}{24\pi
b^{4}}(c_{1}\Lambda'+c_{2}\Lambda)+\frac{\eta^{2}\Lambda'}{3}P_{b}(a,b,b)-
\frac{1}{8640\pi^{2}}(\Lambda'^{2}-\Lambda^{2}).
\end{equation}
In $b\rightarrow \infty$the outer sphere disappears. In this case
one can see from large $z$ behavier of $I_{\nu}(z)$ and
$K_{\nu}(z)$, that interaction force $P_{a}(a,b,a)$ vanishes.
Therefore $P^{tot}(a)$ is exactly the result of Ref. \cite{set2}
for total pressure on spherical shell with zero thickness.\\
In another limiting case, when $a,b\rightarrow$ and $b-a=l$, the
system of two concentric sphere transform to the two parallel
plates configuration with distance $l$. In this limit the first
term in Eqs. (31) and (32) vanishes. Using asymtotic formula for
Bessel function one can show that interaction forces in Eqs. (31)
and (32) change as standard result for the Casimir forces acting
on parallel plates configuration (see section 10 of Ref
\cite{Sahar}).
 These total pressures may be both negative or positive. To see the
different possible cases, let us first assume $\Lambda
> \Lambda'$, noting that $c_{1}>c_{2}$, then
\begin{equation}\label{cond1}
c_{1}\Lambda+c_{2}\Lambda'>0,
\end{equation}
therefore $\frac{\bar F(a)}{A}>0 $. Now noting that
$P_{a}(a,b,a)<0$, second term in Eq. (\ref{ptot1}) is always
positive, therefore the Casimir force on the inner shell is
repulsive, but in this case the gravitational part $P^{(g)}(a)$
is negative. Therefore the total pressure $P^{tot}(a)$ may be
either negative or positive. Given $P^{tot}(a)>0$ initially, then
the initial expansion of the shell leads to a change of the
Casimir part of the pressure. This change, depending on the
details of the dynamics of the shell, may be an increase or a
decrease. Therefore,the initial expansion of the shell may end and
a contraction phase begins. Given $P^{tot}(a)<0$, there is an
initial contraction which ends up at a minimum radius. For the
case $\Lambda>\Lambda'$ situation of the outer sphere is as follows:\\
Let us first assume
\begin{equation}\label{cond2}
c_{1}\Lambda'+c_{2}\Lambda >0,
\end{equation}
then$\frac{\bar F(b)}{B}> 0$, the gravitational part $P^{(g)}(b)$
is positive also, second term is negative. Therefore the total
pressure $P^{tot}(b)$ may be either negative or positive. In this
case  situation is similar to the inner sphere. Now consider the
case
\begin{equation}\label{cond3}
c_{1}\Lambda'+c_{2}\Lambda <0.
\end{equation}
In this case $\frac{\bar F(b)}{B}< 0$, the gravitational part is
positive, interaction term is negative  and above discussion for
recent situation is also correct.\\
Now consider the case $\Lambda < \Lambda'$ and also
\begin{equation}\label{cond4}
c_{1}\Lambda+c_{2}\Lambda' >0.
\end{equation}
Similar to the previous for inner shell Casimir force is
repulsive, the gravitational part is also repulsive, then the
total pressure $P^{tot}(a)$ is always positive. Therefore the
inner shell expands without any limitation.\\
Let us assume
\begin{equation}\label{cond5}
c_{1}\Lambda+c_{2}\Lambda' <0,
\end{equation}
then $\frac{\bar F(a)}{A}< 0$, therefore the total pressure
$P^{tot}(a)$ may be either negative or positive, and all cases of
contraction, expansion may be appear.\\
Now we consider the situation of outer sphere, noting that
$\Lambda < \Lambda'$, then $\frac{ \bar F(b)}{B}$ is positive,
interaction force and the gravitational part are negative. Then
the total pressure $P^{tot}(b)$ may be repulsive or contractive.

    \section{Conclusion}
    We have considered two concentric spherical shells in de Sitter
    space with a massless scalar field, coupled conformally
    to the background, satisfying the Dirichlet boundary
    conditions. Our calculation shows that interaction force
    between two spheres in de Sitter space, similar to the flat
    space is negative, therefore, interaction forces between two
    spheres are attractive, similar to the parallel plate
    configuration. There is another similarity between interaction
    force in concentric spheres and boundary part force acting on
    parallel plates in de Sitter space, as one can see in our
    previous work \cite{set3}, the boundary part pressure acting on
    plates depends only on the cosmological constant between
    the plates, which is like the case of interaction force between
    concentric spheres that as well depends on the cosmological
    constant in spherical layer region.\\The final result
    for total pressure, which in this paper has been obtained, in
    the limit $b\rightarrow \infty$, corresponds to the result of
    \cite{set2} for spherical shell with zero thickness. In the
    limit $a,b\rightarrow \infty$ and $b-a=l$ corresponds to the
    result of \cite{set3} for parallel plates configuration with
    distance $l$.
     Total stress, which
    acts on the each single spheres shows that the detail dynamics
    of spherical symmetry domain wall depends on different
    parameters, and all cases of contraction and expansion may
    appear.
  \vspace{3mm}

{\large {\bf  Acknowledgement }}
 \vspace{1mm}
\small.

 We would like to thank Mr.A.Rezakhani for reading
manuscript.


\begin{thebibliography}{99}

 \bibitem {mueller}
 G. Plunien, B. Mueller, W. Greiner, Phys. Rep. 134, 87(1986)
 \bibitem {trunov} V. M. Mostepaneko, N. N. Trunov, Sov. Phys. Usp. 31(11)
 November 1988
 \bibitem{Casimir}H. B. G. Casimir, Proc. K. Ned. Akad. Wet. 51, 793(1948)
 \bibitem{Remeo}E. Elizalde, S. D. Odintsov, A. Romeo, A. A. Bytsenko and
 S. Zerbini, Zeta Regularization Techniques with Applications(World
 Scientific, Singapore, 1994)
 \bibitem{Elizalde}E. Elizalde, Ten Physical Applications of
 Spectral Zeta Functions, Lecture Notes in
 Physics(Springer Verlag, Berlin, 1995)
 \bibitem{Moste}V. M. Mostepanenko and N. N. Trunov. The Casimir Effect
 and its Applications. (Oxford Science Publications, New York,
 1997)
 \bibitem{Sahar}A. A. Saharian. The Generalized Abel-Plana
 Formula, Applications to Bessel Functions and Casimir
 Effect. hep-th/0002239
 \bibitem{Birrell}N. D. Birrell and P. C. W. Davies, Quantum Fields in
 Curved Space, (Cambridge University Press, 1986)
 \bibitem{Boyer}T. H. Boyer, Phys. Rev. 174, 1764(1968)
 \bibitem{Schw}K. A. Milton, L. L. DeRaad, and
 J. Schwinger, Ann. Phys. (N.Y)115, 338(1978)
 \bibitem{Nester}V. V. Nesterenko, I. G. Pirozhenko, Phys. Rev. D57, 1284(1998)
 \bibitem{Milton}C. M. Bender, K. A. Milton, Phys. Rev. D50, 6547(1994)
 \bibitem{Mil}K. A. Milton. Phys. Rev. D55, 4940(1997)
 \bibitem{Cog}G. Cognola, E. Elizalde, K. Kirsten. Casimir Energies for
 Spherically Symmetric Cavities. hep-th/9906228
 \bibitem{bek1}M. Bordag, E. Elizalde and K. Kirsten, J. Math.
 Phys {\bf 37}, 895(1996).
 \bibitem{bek2}M. Bordag, E. Elizalde, B. Geyer and K. Kirsten,
 Comm. Math. Phys. {\bf 179}, 215 (1996).
 \bibitem{Bord}M. Bordag, E. Elizalde, K. Kirsten and S. Leseduarte.
 Phys. Rev. {\bf D56}, 4896 (1997).
 \bibitem{Eli2}E. Elizalde, M. Bordag and K. Kirsten. J. Phys. A:
 Math. Gen. {\bf 31}, 1743 (1998).
  \bibitem{Bay}S. Bayin and M. Ozcan, Phys. Rev. D48, 2806(1993)
 \bibitem{Bay2}S. Bayin andM. Ozcan, J. Math. Phys. 38, 5240(1997)
 \bibitem{Set}M. R. Setare and A. A. Saharian. Int. J. Mod. Phys. {\bf
 A16}, 1463(2001).
 \bibitem{set1}M. R. Setare and A. H. Rezaeian. Mod. Phys. Lett.{\bf A15}, 2159
 (2000).
 \bibitem{Ant}F. Antonsen and K. Borman. Casimir Driven Evolution of
 the Universe. gr-qc/9802013
 \bibitem{Sahar1}A. A. Saharian. Phys. Rev. {\bf D63}, 125007,
 (2001).
 \bibitem{set2}M. R. Setare and R. Mansouri. Casimir effect for
 spherical shell in de Sitter space. gr-qc/0010028. To appear in
 Class and Quant. Grav.
 \bibitem{set3}M. R. Setare and R. Mansouri. Casimir stress on
 parallel plates in de Sitter space. hep-th/0104160. To appear in
 Class and Quant. Grav.
  \bibitem{Dowk}J. S. Dowker, R. Critchley, Phys. Rev. D13, 3224(1976)





\end{thebibliography}
\end{document}